\documentclass[pra,twocolumn,showpacs,amsmath,amssymb,superscriptaddress]{revtex4}
\usepackage[english]{babel}
\usepackage{epsfig}
\usepackage{color}
\usepackage{graphicx}

\emergencystretch=\maxdimen
\begin{document}

\title{Bosonic Kondo-Hubbard model}
\author{T. Flottat}
\affiliation{
INLN, Universit\'e de Nice-Sophia Antipolis, CNRS;
1361 route des Lucioles, 06560 Valbonne, France.
}
\author{F. H\'ebert}
\affiliation{
INLN, Universit\'e de Nice-Sophia Antipolis, CNRS;
1361 route des Lucioles, 06560 Valbonne, France.
}
\author{V. G. Rousseau}
\affiliation{
Department of Physics and Astronomy, Louisiana State University,
  B\^aton Rouge, Louisiana 70803, USA.
}

\author{R. T. Scalettar}
\affiliation{
Physics Department, University of California, Davis, California 95616, USA.
}

\author{G. G. Batrouni}
\affiliation{
INLN, Universit\'e de Nice-Sophia Antipolis, CNRS;
1361 route des Lucioles, 06560 Valbonne, France.
}
\affiliation{
Institut Universitaire de France, 103 boulevard Saint Michel, 75005 Paris, France.
}

\begin{abstract}
  We study, using quantum Monte-Carlo simulations, the bosonic
  Kondo-Hubbard model in a two dimensional square lattice.  We explore
  the phase diagram and analyse the mobility of particles and magnetic
  properties. At unit filling, the transition from a
    paramagnetic Mott insulator to a ferromagnetic superfluid appears
    continuous, contrary to what was predicted with mean field.  For
  double occupation per site, both the Mott insulating and superfluid
  phases are ferromagnetic and the transition is still continuous.
  Multiband tight binding Hamiltonians can be realized in optical
  lattice experiments, which offer not only the possibility of tuning
  the different energy scales over wide ranges, but also the option of
  loading the system with either fermionic or bosonic atoms.
\end{abstract}


\maketitle

\section{Introduction}

In condensed matter systems, the interaction between mobile particles
and fixed magnetic impurities, known as Kondo physics, has been a very
important topic for the last 50 years.  From the original explanation
of the resistance minimum in metals with magnetic impurities by Kondo,
to the investigation of the properties of heavy fermion materials,
the interaction between particles and localized spins has revealed a
variety of interesting physical phenomena
\cite{hewson93,tsunetsugu97}.
Indeed, the competition between magnetic ordering and singlet formation
in Kondo and related materials has offered some of the most fundamental
examples of quantum phase transitions \cite{wang06}, and investigations
of the effects of interplay of the distinct spin and particle
contributions to the susceptibility, and of
dilution\cite{assaad02,seo14}, are at the frontier of  the investigation
of many materials, including the `115' heavy fermion family
\cite{shirer12,seo14}

In addition to these solid state systems,
with the recent experimental advances in ultracold atomic physics, it
is now possible to build systems of atoms on optical lattices with
atoms occupying different bands
\cite{hemmerich11,mueller07,clement09}. This opens the possibility to
use the atoms located in the lowest band as localized particles,
magnetic centres, which will interact with mobile particles located in
higher energy bands.  Such systems would be analogues of Kondo
problems but with the possibility to use bosonic particles instead of
fermionic ones, systems that have not been extensively studied and are
not available in condensed matter physics.

We will study here a system similar to the Kondo-Hubbard lattice
problem \cite{tsunetsugu97,feldbacher02,yanagisawa95,fazekas97} with
interacting spin 1/2 bosons instead of fermions. Mobile bosons
are free to move on the lattice and interact repulsively on
site.  In addition, there is an antiferromagnetic (AF) coupling to an
ensemble of spin 1/2 magnetic centres, one for each site of the
lattice.  This model was introduced by Duan \cite{duan04}
to describe the following system: the localized bosonic species (`spins') are atoms in the lowest band of
an optical lattice with a potential barrier which prohibits tunneling.
The mobile species occupy an upper band, to which they have been excited
through the applicaton of periodic Raman pulses, which allows
tunneling. 
This model was 
studied in detail with different analytical techniques by
Foss-Feig and Rey \cite{fossfeig11}. In \cite{fossfeig11}, exact
results were derived for the small and large interaction limits and
the intermediate regime was studied using mean-field theory, for
different densities of particles.  For one mobile particle per site,
they observed a first order transition between a Mott insulator (MI) phase
and a superfluid (SF) phase as the interaction is lowered. In the Mott
phase, there are singlets of bosons and spin and no long range
magnetic order whereas the SF phase shows long range ferromagnetic
(FM) order for the bosons and the spins.  For two or more particles
per site, the Mott phase is already ferromagnetic and the transition
to the ferromagnetic superfluid state is continuous.

In this paper, we will use exact quantum Monte Carlo simulations to
study this bosonic Kondo-Hubbard model \cite{duan04} and determine
exactly the phase diagram and magnetic properties of the system at
zero and finite temperatures $T$ and compare with results previously
obtained with mean-field approximations \cite{duan04,fossfeig11}.  In
Sec. II, we introduce the model, the numerical technique we used and
the quantities we will measure to characterize the phases.  In
Sec. III, we study the transport properties and Green functions of the
system to draw its phase diagram at $T=0$.  In Sec. IV we analyse in
more detail the nature of the quantum phase transitions.  Sec. V is
devoted to a careful analysis of the magnetic properties in the
ground state and Sec. VI presents the evolution of the phases observed
at $T=0$ as the temperature is increased.

\section{Bosonic Kondo-Hubbard model}

\begin{figure}
\includegraphics[width=6.5cm]{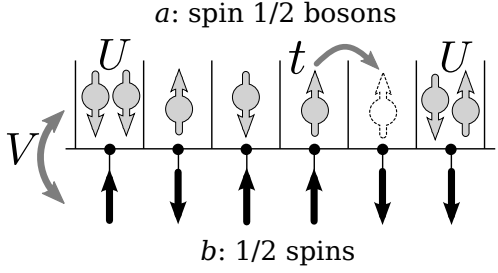}
\caption{Schematic representation of the model. Spin-$1/2$ bosons
  (upper layer $a$) are moving through the lattice with hopping $t$
  and are subject to on-site repulsion $U$.  They interact
  antiferromagnetically with strength $V$ with the lower spin layer
  $b$. The lattice represented here is a 1D chain; the studied system
  is a 2D square lattice.
\label{fig:model}
}
\end{figure}

The system we consider includes two types of objects that are coupled
antiferromagnetically (AF): spin-$1/2$ bosons which hop on a 2D square
lattice and, on every site, a fixed spin-$1/2$ magnetic impurity. In
the following, we will denote the bosons as particles of type $a$ and
the fixed spins as particles of type $b$. An easy way to visualize
this system is to use two layers, $a$ and $b$, one filled with bosons,
the other with spins (Fig. \ref{fig:model}).  The Hamiltonian reads
\begin{eqnarray}
{\cal H}  &=&
-t \sum_{\langle i,j\rangle,\sigma}\left(a^\dagger_{i,
  \sigma}a^{\phantom{\dagger}}_{j, \sigma} + {\rm
  h.c.}\right)\label{hopping}\\ 
&+&\frac{U}{2} \sum_{i} n_{a, i}\left(n_{a, i} - 1
\right) \label{repulsion}\\ 
&+& V \sum_{i}{\bf S}_{a, i}\cdot{\bf S}_{b,
  i}\label{exchange} 
\end{eqnarray}
The operators $a^\dagger_{i,\sigma}$ and $a_{i,\sigma}$ create or
destroy an $a$-type boson of spin $\sigma$ on site $i$.  The system is
a 2D square lattice with $L^2$ sites where $L$ is the length of the
lattice. There is always one $b$ spin per site and we will vary the
number $N_a$ of $a$ bosons.  The Hamiltonian includes a hopping term
(Eq. \ref{hopping}) and an on site repulsion (Eq. \ref{repulsion}) for
the $a$ bosons.  $n_{a,i} = \sum_\sigma n_{a,i,\sigma} = \sum_\sigma
a^\dagger_{i,\sigma} a^{\phantom{\dagger}}_{i,\sigma}$ is the total
number of bosons on site $i$.  The hopping parameter $t$ sets the
energy scale and the on-site repulsion energy is $U$.

The last term (Eq. \ref{exchange}) of the Hamiltonian is an
antiferromagnetic coupling between the boson magnetic moment and the
fixed $b$ spins.  ${\bf S}_{a,i} = (S^x_{a,i}, S^y_{a,i},S^z_{a,i})$
gives the spin of the bosons and its components are given by
\begin{equation}
S^\alpha_{a,i} = \sum_{\sigma,\sigma'} a^\dagger_{i,\sigma} S^\alpha_{\sigma,\sigma'} a^{\phantom{\dagger}}_{i,\sigma'}
\end{equation}
where the $S^\alpha_{\sigma,\sigma'}$ are the three standard spin 1/2
matrices.  The $b$ spins are also described by these matrices ${\bf
  S}_{b,i} = (S^x_{b,i}, S^y_{b,i},S^z_{b,i})$.  To specify a state of
the system one should give the state $S^z_{b,i} = \pm 1/2$ of each $b$
spin and the number $n_{a,i,\sigma}$ of up or down $a$ bosons present
on each site $i$.  The conventional quantum number $s$ will be used
when discussing the possible eigenvalues of angular momentum ${\bf
  S}^2 = s(s+1)$.

The last term of the Hamiltonian is the Kondo interaction, here in the
form used in the Kondo insulators where the moving particles interact
with a network of magnetic moments. This is different from the
original Kondo problem where the moving particles are coupled to a
small number of magnetic ``impurities" distributed randomly
\cite{hewson93} and more similar to the ``Kondo lattice"
\cite{fazekas91}. 
Other differences with the original Kondo problem are
that our moving particles are not free but interacting with each other
and, of course, they are bosons and not fermions. 
Studying an equivalent spin 1 model would be interesting but is more demanding numerically.
The spin 1/2 model allows us
to compare with the results from \cite{fossfeig11,duan04},
which also propose an experimental realization of the model. Furthermore
the qualitative physics should not be different with a spin 1 model.

To study this system, we used the quantum Monte Carlo SGF algorithm
\cite{rousseau08,rousseau08-2} that allows exact calculations of
physical observables at finite temperature on clusters of finite size
(up to $L=14$).  We are especially interested in one and two-body
Green functions that are possible to calculate with the SGF algorithm.
To extract the properties of the ground state, we used large inverse
temperatures $\beta = 1/kT$, up to $\beta t = 25$.

We studied the one-body Green functions for the bosons
$G_{a,\sigma}(R)$,
\begin{equation}
G_{a,{\sigma}}(R) = \frac{1}{2L^2} \sum_i \langle a^\dagger_{i,\sigma}
a^{\phantom{\dagger}}_{i+R,\sigma} + a^\dagger_{i+R,\sigma}
a^{\phantom{\dagger}}_{i,\sigma} \rangle.
\end{equation}
The condensed fraction $\rho(k=0)$ is the Fourier transform at $k=0$,
$\rho(k=0) = \sum_{R,\sigma} G_{a,{\sigma}}(R)/L^2$.  The superfluid
density $\rho_s$ can be measured using the standard relation with
fluctuations of the winding number $\rho_s = \langle W^2 \rangle /
4t\beta$ as the total number of bosons is conserved \cite{ceperley89}.

We also studied anticorrelated two-body Green functions which describe
exchange of particles or spins at long distances, which then correspond to
opposite, anticorrelated movements. They are generally
important for multispecies Hamiltonians with repulsive interactions
where exchanges are the dominant effects in the strongly interacting regimes
\cite{kuklov03}. In this case, they are conveniently expressed in
terms of spin degrees of freedom
\begin{eqnarray}
G_{aa}(R) &=&\frac{1}{2L^2} \sum_i \langle
a^\dagger_{i,\uparrow}a^{\phantom{\dagger}}_{i,\downarrow}
a^{\phantom{\dagger}}_{i+R,\uparrow} a^\dagger_{i+R,\downarrow} +{\rm
  h.c.}\rangle \nonumber \\ 
&=&\frac{1}{2L^2} \sum_i\langle S^+_{a,i}
S^-_{a,i+R} + {\rm h.c.}\rangle\\ 
G_{bb}(R) &=&\frac{1}{2L^2}
\sum_i\langle S^+_{b,i} S^-_{b,i+R} + {\rm h.c.}\rangle\\ 
G_{ab}(R)&=&\frac{1}{2L^2} \sum_i \langle
a^\dagger_{i,\uparrow}a^{\phantom{\dagger}}_{i,\downarrow}
S^-_{b,i+R}+{\rm h.c.}\rangle \nonumber \\ 
&=&\frac{1}{2L^2}
\sum_i\langle S^+_{a,i} S^-_{b,i+R} + {\rm h.c.}\rangle
\end{eqnarray}
As $(S^+_iS^-_{i+R}+S^-_iS^+_{i+R}) = 2 (S^x_{i}S^x_{i+R} +
S^y_{i}S^y_{i+R})$, they correspond to the spin correlations in the
$x-y$ plane.

Adding the
spin-spin correlations along the $z$ axis (which are diagonal
quantities) to the correlations in the $xy$ plane that were obtained
through Green functions, we obtain the complete spin-spin correlations.
For example,
\begin{equation}
S_{aa}(R) = \frac{1}{L^2} \sum_i \langle {\bf S}_{a,i}\cdot {\bf
  S}_{a,i+R}\rangle.
\end{equation}
Similar definitions hold for correlations $S_{bb}(R)$ between the $b$
spins and for correlations $S_{ab}(R)$ between bosons and spins.  We
will denote by ${\bf S}_{\rm tot} = \sum_i ({\bf S}_{a,i} + {\bf
  S}_{b,i})$ the total spin, or total magnetization, of the system,
which is expressed as a sum of spin correlations functions
$$ S^2_{\rm tot} = \sum_{R} \left[S_{aa}(R) + 2 S_{ab}(R) +
  S_{bb}(R)\right]
$$


\section{Phase Diagram}

\begin{figure}[h]
\includegraphics[width=8.5cm]{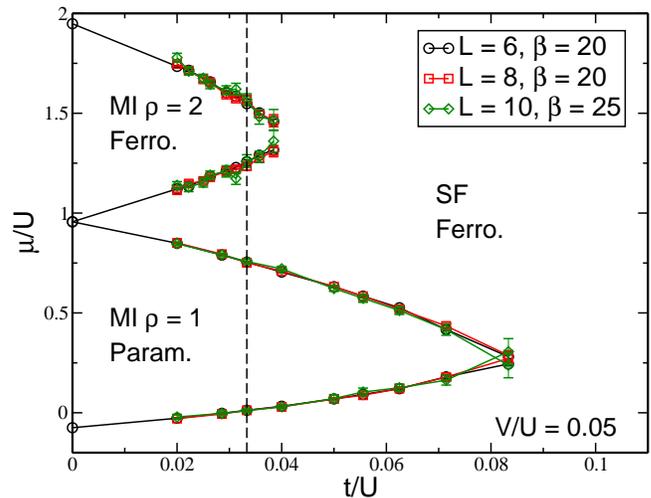}
\caption{Phase diagram for several system sizes $L$. The insulating
  phase at $\rho = 1$ is paramagnetic while the other phases are
  ferromagnetic. The dashed line indicates the cut in Figure
  \ref{cutrhosf}.
  \label{diagphase}}
\end{figure}

We first show the phase diagram in the $(t/U, \mu/U)$ plane at $T=0$
for a fixed value of $V/U =0.05$ (Fig. \ref{diagphase}).  
For small values of $V/U$, some of the transitions in
the system are predicted to be first order \cite{fossfeig11}.  At $T=0$, using a
canonical simulation, the chemical potential is given by the energy
difference $\mu(N_a) = E(N_a+1) - E(N_a)$ where $N_a$ is the number of
bosons. This allows to draw the boundaries of the $n$th Mott lobe with
$n$ bosons of type $a$ per site by measuring the energy of the system
with $N_a = n L^2, n L^2+1$, and $n L^2-1$ particles. In the $t/U=0$
limit, an analytical calculation yields \cite{fossfeig11} $\mu_{0
  \rightarrow 1} = -3V/4$, $\mu_{1\rightarrow 2} = U - V/4$,
$\mu_{2\rightarrow 3} = 2U - V/4$, where $\mu_{n\rightarrow n+1}$ is
the value at which the density changes from $n$ to $n+1$. In
Fig. \ref{diagphase}, we exhibit the $\rho=1$ and $\rho=2$ insulating
Mott lobes.  Outside of these Mott lobes, the system is superfluid and
Bose condensed, as we will show below. The first Mott lobe is
paramagnetic and the rest of the phase diagram has ferromagnetic
correlations of the bosons.

Compared to other studies of spin-$1/2$ bosons or mixtures of
particles \cite{deforges11}, the Mott lobes are
larger. As expected, the presence of the Kondo interaction favours
insulating behaviour: The tip of the $\rho=1$ Mott lobe is located
around $t/U \simeq 0.08$ for $V/U = 0.05$ whereas it is located around
$t/U\simeq 0.06$ for $V=0$. As $V/U$ is increased up to 0.25, the tip
shifts further to $t/U \simeq 0.10$ (see Fig. \ref{SVU.25rho1}).  This
robustness is not surprising since, for $\rho=1$, the Mott gap is
equal to $U+V/2$ \cite{fossfeig11} in the $t=0$ limit.

\begin{figure}[h]
\includegraphics[width=8.5cm]{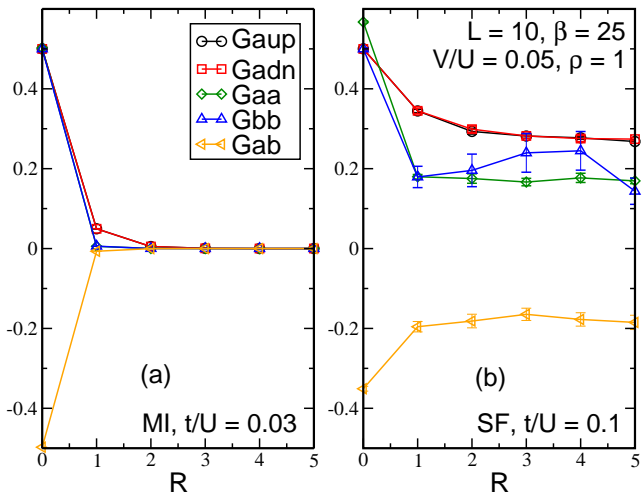}
\caption{The one-body and the anticorrelated Green functions versus
  distance $R$ in the Mott insulator (a) and superfluid phase (b) for
  $\rho = 1$. (a) all movement is suppressed as the particles form
  singlets. (b) movement of individual particles as well as
  anticorrelated exchanges are observed.
\label{green1}}
\end{figure}

Analysing the Green functions at $\rho=1$, we observe that they all
decrease rapidly to zero with distance in the Mott phase and there is
no phase coherence (Fig.\ref{green1} (a)).  This is expected for the
one-body Green functions but is also the case for the anticorrelated
Green functions.  In the $t=0$ limit, to minimize the AF energy
between the spins and bosons, the magnetic moment of the boson forms a
singlet with the spin located at the same site.  As these singlets are
formed, there is a unique Mott state in the $t=0$ limit and there is
no possibility to exchange bosons of different spins
\cite{kuklov03}. This behaviour is probably maintained throughout the
$\rho=1$ Mott lobe, even at $t\ne 0$.  This is verified by the value
of $G_{ab}(0) \simeq -0.5$ which shows the on-site antiferromagnetic
correlation of the magnetic moments and will be confirmed below by
direct measurements of the magnetic correlations.

\begin{figure}[h]
\centerline{\includegraphics[width=6.5cm]{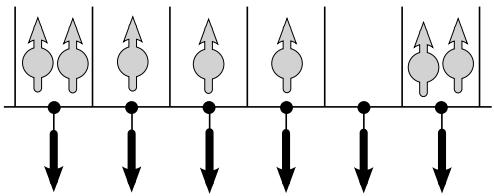}}
\caption{The magnetic order that appears in the system. While the
  bosons and the spins form ferromagnetic layers, these two layers are
  coupled antiferromagnetically. \label{magorder}}
\end{figure}

In the superfluid phase at $\rho=1$ (Fig.\ref{green1} (b)), on the
other hand, all the Green functions show long range order.  The long
range order of the one-body Green function $G_{a,\sigma}$ shows that
the system is Bose condensed. The non zero value of $G_{aa}$ shows
that, in addition to the individual movement of the particles,
exchange moves are important degrees of freedom for this phase.
$G_{bb}$ is non zero, which shows that the spins are correlated. This
correlation is mediated by the movement of the bosons as the spins are
not directly linked to each other. This is confirmed by the
observation that the boson-spin Green function, $G_{ab}$, is also non
zero. While $G_{aa}$ and $G_{bb}$ are positive, which signals
ferromagnetic behaviour, $G_{ab}$ is negative which is expected since
the coupling between bosons and spins is AF.  The picture that emerges
from these results is that the bosons and the spins form ferromagnetic
phases but that these two species are coupled in an antiferromagnetic
way (Fig.\ref{magorder}).


\begin{figure}[h]
\includegraphics[width=8.5cm]{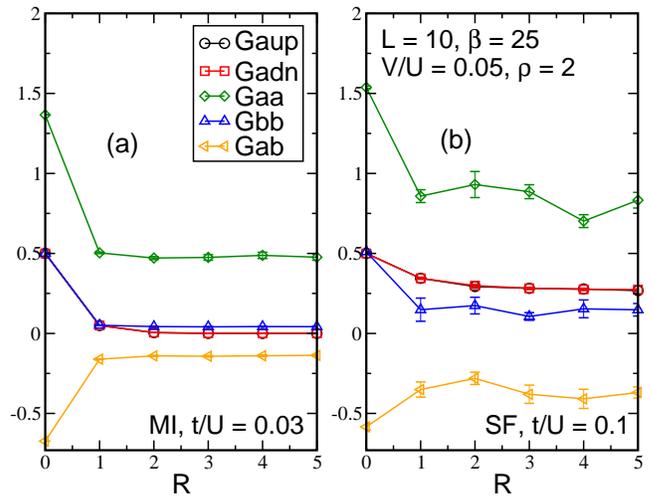}
\caption{The one-body and the anticorrelated Green functions versus
  distance $R$ in the Mott insulator (a) and superfluid (b) phases for
  $\rho = 2$.  (a) Individual movement are suppressed but
  anticorrelated moves are possible in the Mott phase. (b) All kind of
  movements are present in the superfluid phase.
\label{green2}}
\end{figure}

For $\rho=2$, the Mott phase behaves differently from $\rho=1$
(Fig. \ref{green2} (a)). The individual movement of particles are
still suppressed: $G_{a,\sigma}$ goes to zero with distance. However
exchange movements are present and, consequently, there are couplings
between the spins as is seen from the anticorrelated Green functions
taking finite values at large distances. This can be understood by
noting that the ground state in the $t=0$ limit is not unique
\cite{fossfeig11}.  This degeneracy will be lifted by a non zero
hopping term and give a ground state with ferromagnetic correlations
between the bosons.  The magnetic order present in this phase is then
similar to the one observed in the superfluid phase.

The $\rho=2$ superfluid phase (Fig. \ref{green2} (b)) shows the same
qualitative behaviour as the $\rho=1$ SF phase. However the dominant
behaviour in this case is the anticorrelated movements of bosons
whereas they were individual movements of bosons at $\rho=1$. This is
expected since, in a strongly correlated system, particles can move
with a partner while individual movement is suppressed. Of course,
as the density increases, the correlations become more prominent.

\section{Quantum Phase transitions\label{secqpt}}

\begin{figure}[h]
\includegraphics[width=8.5cm]{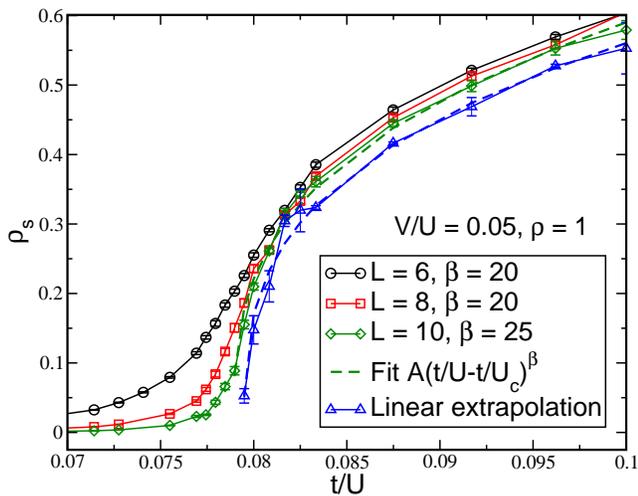}
\caption{Superfluid density, $\rho_{s}$, versus $t/U$ in the first
  Mott lobe. We observe a  continuous behavior for $\rho_{s}$. The blue curve
with triangles
represents a linear extrapolation to $L=\infty$ of the results obtained for the three sizes
$L=6,\,8,\,10$. The curves for $L=10$ and the extrapolated one
have been fitted with a power law behavior $\rho_s \propto (t/U - t/U_c)^{\beta}$
(dashed lines). In both cases, we found $t/U_c \simeq 0.79$ and $\beta \simeq 0.33$.
\label{rhosf1}}
\end{figure}

\begin{figure}[h]
\includegraphics[width=8.5cm]{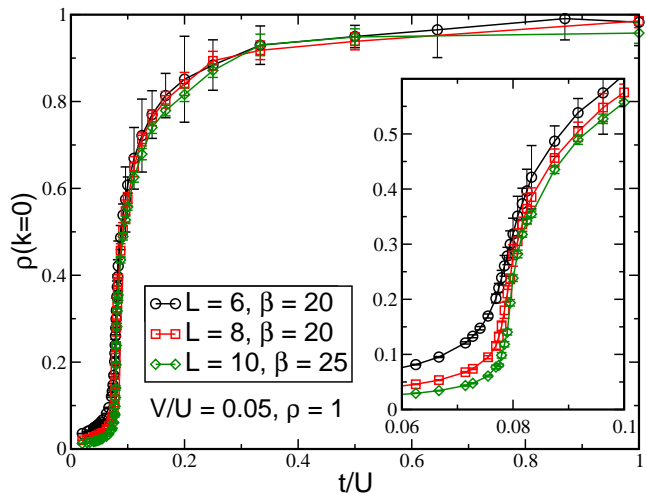}
\caption{Condensate fraction, $\rho(k=0)$, versus $t/U$ in the first
  Mott lobe. The shape is very similar to figure \ref{rhosf1},
  confirming the order of the transition.
\label{rhosfk1}}
\end{figure}

\begin{figure}[h]
\includegraphics[width=8.5cm]{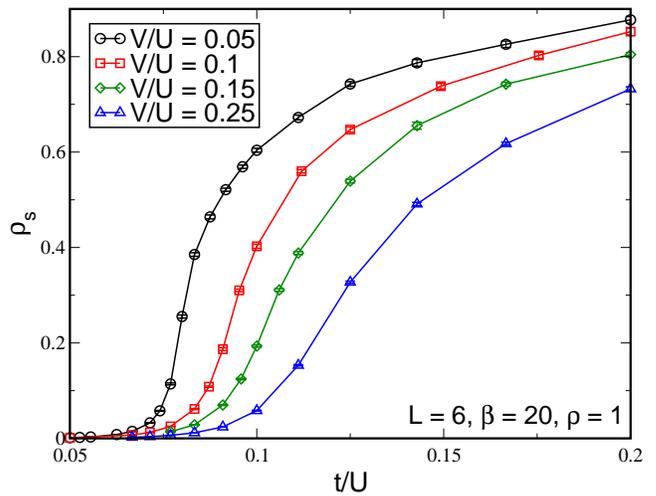}
\caption{Superfluid density, $\rho_{s}$, versus $t/U$ in the first Mott
  lobe, for different ratio $V/U$. As we increase this ratio,
  $\rho_{s}$ becomes smoother, compared to  Fig.~\ref{rhosf1}
and the transition remains second order.
\label{compVrhosf1}}
\end{figure}

We now analyse the nature of the phase transitions
  between the Mott and the superfluid phases by examining the
  behaviour of the superfluid density. In the Mott lobe $\rho_s=0$; as
  $t/U$ is increased at fixed $\rho=1$ and $V/U=0.05$, we observe a
  seemingly continuous transition from the MI to the SF
  (Fig. \ref{rhosf1}).  We fit the curves near the transition with a
  form $\rho_s \propto (t/U - t/U_c)^{\beta}$ (Fig. \ref{rhosf1}) and
  found $\beta \simeq 0.33$. It is however difficult to distinguish a
  continuous transition with a small $\beta$ coefficient (which gives
  a large slope close to the transition) from a discontinuous one on
  such finite size systems.  The condensate density $\rho(k=0)$ shows
  a similar behaviour (Fig. \ref{rhosfk1}), as expected in two
  dimensions in the zero temperature limit where both $\rho_s$ and
  $\rho(k=0)$ are non zero in the superfluid phase.  A continuous
  transition is in disagreement with the MF analysis from
  \cite{fossfeig11} that predicts a first order, discontinuous,
  transition when $V/U< 0.1$, which is the case here.
  To confirm this result, we calculated $\rho(\mu)$ in the canonical
  ensemble and did not find a negative compressibility region which
  would have been a clear sign of a discontinuous transition
  \cite{batrouni00}.  We also observe no sign of a discontinuity at
  the tip of the $\rho=1$ Mott lobe, if there is one, as predicted in
  \cite{fossfeig11}, it is too small to discern for the system sizes
  accessible to us. In Fig. \ref{compVrhosf1}, we show the evolution
  of this behaviour for a given size as $t/U$ is increased.  As $V/U$
  becomes larger the behaviour becomes smoother and the transition
  still appears continuous.

\begin{figure}
\includegraphics[width=8.5cm]{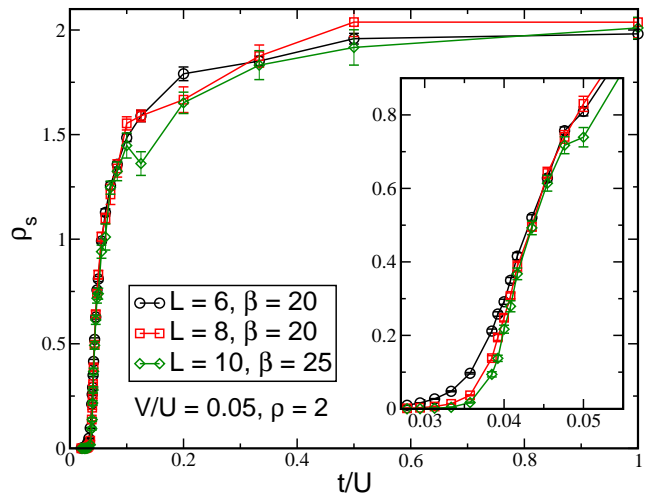}
\caption{Superfluid density, $\rho_{s}$, versus $t/U$ in the second
  Mott lobe. As for the $\rho=1$ case (Fig. \ref{rhosf1}), we do not
  observe any discontinuity in $\rho_{s}$ and the transition is
  second-order.
\label{rhosf2}}
\end{figure}

In Fig. \ref{rhosf2} we show $\rho_s$ as a function of $t/U$ for the
$\rho=2$ Mott-SF transition and do not observe a discontinuity between
the two phases: the transition is continuous for all $V/U$.

\begin{figure}[h]
\includegraphics[width=8.5cm]{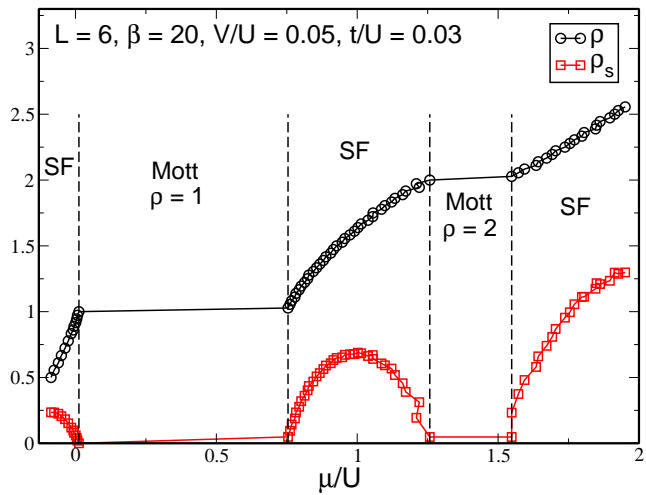}
\caption{Cut of figure \ref{diagphase} for $t/U = 0.03$, showing
  the total density, $\rho$, and the superfluid density, $\rho_{s}$,
  as functions of $\mu/U$. All the transitions are continuous. 
\label{cutrhosf}}
\end{figure}

In figure \ref{cutrhosf}, we examine the dependence of $\rho$ and
$\rho_s$ on the chemical potential $\mu$ along the dashed line in
Fig. \ref{diagphase}. We observe the conventional incompressible Mott
plateaux where $d\rho/d\mu=0$. The superfluid density goes to zero
continuously as these plateaux are approached, showing that the
transition for the $\rho=1$ Mott is, as mentioned earlier, continuous.
$\rho(\mu)$ is also continuous and has a positive slope, there is thus
no sign of a phase separation close to the transition.

\section{Magnetic properties}

We now study in more detail the magnetic properties of the system.  We
plot $S_{\alpha\beta}(L/2)$ at the largest
possible distance $L/2$. For a sufficiently large system this
converges to the square of the magnetization of a given layer. We also
study the on-site correlation to see if singlets are formed between
the $a$ and $b$ layers, and the total spin of the system $S^2_{\rm
  tot}/L^4$.

\begin{figure}[h]
\includegraphics[width=8.5cm]{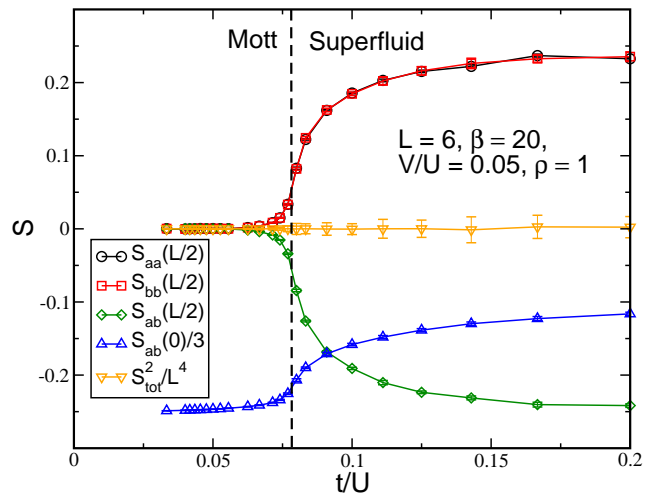}
\caption{Magnetic correlations versus $t/U$ for $\rho=1$ and
  $V/U=0.05$.  The dashed line marks the transition between the Mott
  and the superfluid phases.  There is no magnetic order in the Mott
  phase where the $a$ and $b$ spins form singlets, as shown by the
  values of $S_{ab}(0)$. The superfluid phase shows the magnetic
  behaviour depicted in Fig.\ref{magorder}: intra-species
  ferromagnetic correlations and inter-species antiferromagnetic
  correlations.  At large $t$, the magnetic correlations tend to their
  maximum values.
\label{SVU.05rho1}}
\end{figure}

In Fig. \ref{SVU.05rho1} we plot these quantities as functions of
$t/U$ for $\rho=1$. We observe that in the Mott phase, the magnetic
correlations are always zero. $S_{ab}(0) = -3/4$ signals the formation
of a singlet. As the $a$ and $b$ particles form a singlet on each
site, the absence of magnetic correlations between sites is
reasonable.  In the superfluid phase, on the other hand, we observe
that we no longer have a singlet phase as $S_{ab}(0)$ departs from the
value observed in the Mott phase. We also observe, as anticipated
earlier, that ferromagnetic correlations develop between the $a$
bosons, between the $b$ spins and antiferromagnetic correlations
persist between the two types of particles (corresponding to the
positive values of $S_{aa}(L/2)$ and $S_{bb}(L/2)$ and the negative
value of $S_{ab}(L/2)$).  Deep in the superfluid, the magnetic
correlations take their maximum possible value $|S_{\alpha\beta}(L/2)|
\rightarrow 1/4$.  It should be remarked that the correlation between
the $b$ spins is mediated by the itinerant $a$ bosons, as there are no
direct connections between the spins themselves.  This is similar to
the coupling between localized spins provided by the RKKY interaction
\cite{ruderman54, kasuya56, yosida57} in fermionic systems, although
it is always ferromagnetic in our case.  Within error bars, we have
$S_{aa}(L/2) = S_{bb}(L/2) = - S_{ab}(L/2)$ and, accordingly, the
value of the total spin $S^2_{\rm tot}/L^4$ is zero. This was
predicted in reference \cite{fossfeig11} in the high and low $t/U$
limits but we see here that this seems to be the case also for
intermediate values. We then have very different magnetic behaviour
(independent singlets in the Mott, magnetic order in the SF) with the
same $S_{\rm tot}$.
One should remarks that, whereas the antiferromagnetic
correlations between bosons and spins are imposed by the Hamiltonian
and were expected, the ferromagnetism of the bosons layer appears
spontaneously. There is no term that directly favors the development
of FM correlations between bosons compared to other spin textures.

\begin{figure}[h]
\includegraphics[width=8.5cm]{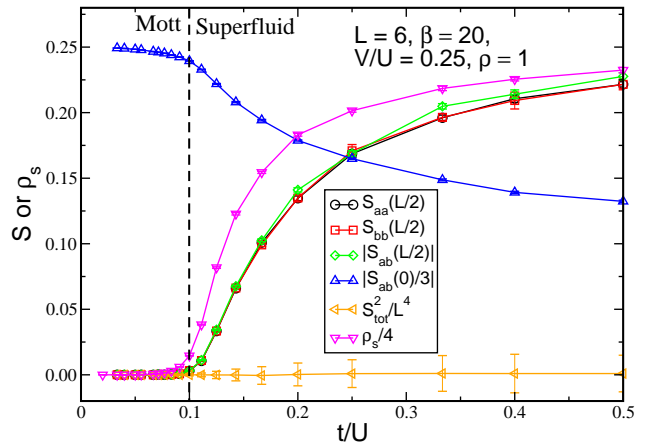}
\caption{Magnetic correlations and $\rho_s$ versus $t/U$ for $\rho=1$
  and $V/U=0.25$.  The physical behaviour is the same as in the
  $V/U=0.05$ case (Fig. \ref{SVU.05rho1}) but it is more difficult to
  establish the magnetic correlations with this larger value of $V$.
\label{SVU.25rho1}}
\end{figure}

Increasing $V$, we observe the same qualitative behaviour with some
quantitative changes (Fig. \ref{SVU.25rho1}). As mentioned earlier,
the Mott-SF transition is shifted towards lower values of $U$ as the
Kondo interaction is added to the repulsion between particles, which
is visible in the behaviour of $\rho_s$. The appearance of the
superfluidity and of the magnetic correlations is once again
simultaneous and corresponds to the disappearance of the spin
singlets.  Finally we remark that the magnetic correlations tend to
their maximum values but that those will be reached for much larger
values of $t/U$. This is understandable as the singlets are more
difficult to break at large $V$ which makes it more difficult to
develop intersite correlations.

\begin{figure}[h]
\includegraphics[width=8.5cm]{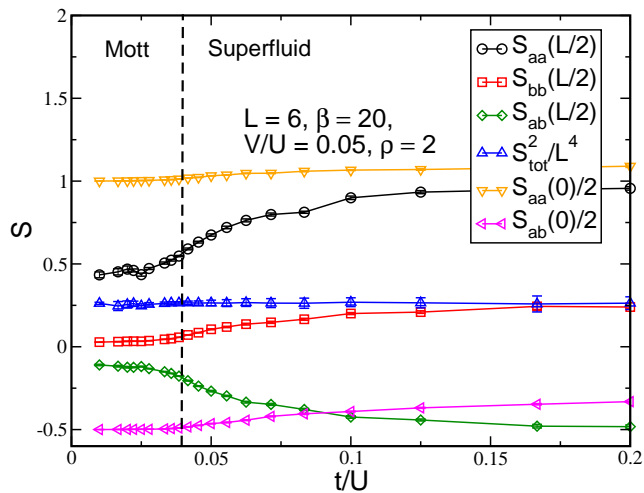}
\caption{Magnetic correlations versus $t/U$ for $\rho=2$ and
  $V/U=0.05$.  The dashed line marks the transition between the Mott
  and the superfluid phases.  The same magnetic order is present in
  the Mott and the superfluid phases although there are two different
  limiting regimes with stronger (for $t/U \gg 1$) or weaker (for $t/U
  \ll 1$) $b$ spin correlations. The magnetic behaviour is the one
  sketched in Fig. \ref{magorder}. The total spin seems constant
  $S^2_{\rm tot}/L^4 \simeq 1/4$.
\label{SVU.05rho2}}
\end{figure}

For $\rho=2$, deep in the Mott phase, the $a$ spins located on the
same site form a total $s_a \approx 1$ moment, which gives $S_{aa}(0)
= s_a(s_a+1) \approx 2$.  This spin is then coupled
antiferromagnetically to a $b$ spin which is shown by the value of
$S_{ab}(0) \approx -1$, giving a total spin-$1/2$ and, consequently,
two degenerate states on each site (Fig. \ref{SVU.05rho2}).  The
kinetic term lifts the degeneracy between these states and we obtain
the magnetic order which is observed in the SF regions and also even
in the Mott lobe. Analytically \cite{fossfeig11}, it was predicted
that $S_{bb}(R\ne 0) = 1/36 \simeq 0.0278$.  For the
  largest value of the interaction used in our simulations $U=100t$,
  we observe $S_{bb}(L/2) = 0.028\pm0.001$ and, as can be seen in
  Fig. \ref{SVU.05rho2}, we reached the regime where $S_{bb}(L/2)$
  saturates at this non zero value.

In the superfluid phase, the system behaves very much as for
$\rho=1$. $S_{aa}(0)$ and $S_{ab}(0)$ depart from their Mott phase
value and increase slightly. $S_{aa}(L/2)$, $S_{bb}(L/2)$, and
$S_{ab}(L/2)$ go to their extreme possible values 1, 1/4 and -1/2,
respectively.

It is predicted \cite{fossfeig11} that $S^2_{\rm tot} = (L^2/2)(L^2/2
+ 1) \approx L^4/4$ in the strong and weak coupling limits.  We
observe that $S^2_{\rm tot}/L^4$ always takes a value compatible with
1/4, for any value of $t/U$. As for $\rho=1$, we observe two different
behaviours for the same common value of $S^2_{\rm tot}/L^4$.

\section{Thermal effects}

\begin{figure}[h]
\includegraphics[width=8.5cm]{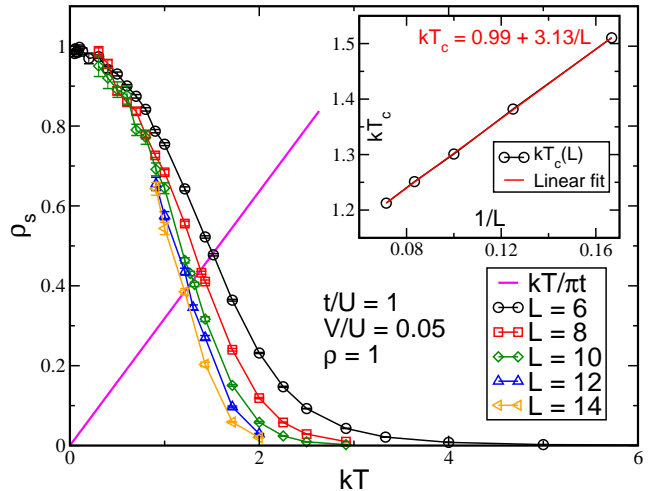}
\caption{Superfluid density, $\rho_s$, versus $T$ for several sizes
  $L$ with $t/U=1$ and $V/U = 0.05$. We calculate the intersection
  between the curves and $kT/\pi t$ to obtain $T_c(L)$. We then
  extrapolate linearly (see inset) the value 
   of $T_c(L)$ to the large size limit to obtain
  the estimate of the transition temperature shown in
  Fig. \ref{diagtemp}.\label{nelson}}
\end{figure}

We analysed the behaviour of the observed phases at finite
temperature.  First we looked at the superfluid density to determine
the extent of the superfluid phase as the temperature is increased.
The thermal phase transition between the superfluid phase and the
normal liquid is of the BKT type.  We performed different finite size
analyses to determine the critical temperature $T_c$ at which $\rho_s$
becomes zero.  First we used linear extrapolations of $\rho_s$ as a
function of $1/L$ for different values of $1/T$.  Then we used Nelson
and Kosterlitz's result \cite{nelson77} $\rho_s(T_c) = kT/\pi t$ to
calculate the temperature $T_c(L)$ at which our curves intersect
$kT/\pi T$ before looking at the $1/L \rightarrow 0$ extrapolation
(Fig. \ref{nelson}).  Finally we used the recently proposed method by
Hsieh {\it et al.}  \cite{hsieh13}.  All three methods gave similar
results for the system sizes accessible to us.

There is no transition between the Mott phase and the normal liquid,
as the Mott phase exists only at zero temperature, strictly
speaking. We calculated the fluctuation of the number of particles
$\kappa = \langle n_{a,i}^2 \rangle - \langle n_{a,i}\rangle^2$ which
exhibits a plateau at small $kT$ before increasing at higher $kT$. We
identify the crossover temperature between the Mott and the liquid
behaviour as the temperature where $\kappa$ departs from this low $T$
value by more than $5\%$. We checked that this definition is valid by
comparing with a measure of $\rho(\mu)$ and finding the $T$ at which
the Mott ``plateaux" disappear in $\rho(\mu)$.

\begin{figure}[h]
\includegraphics[width=8.5cm]{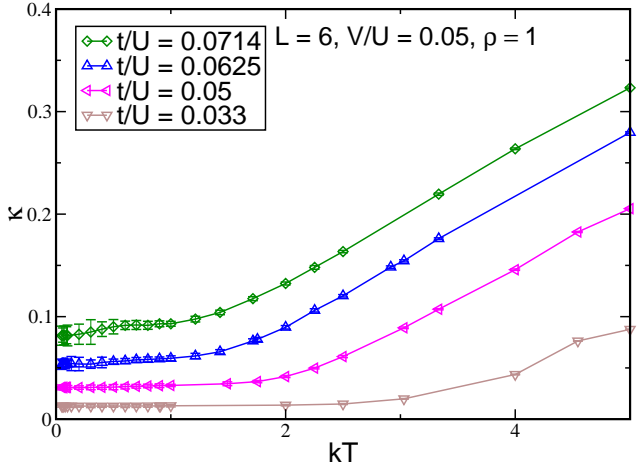}
\caption{The local density fluctuation $\kappa=\langle n_{a,i}^2
  \rangle - \langle n_{a,i}\rangle^2$ versus $kT$ for different
  temperatures. We see that $\kappa$ is almost constant at low $kT$ in
  the Mott region.  We defined the crossover temperature between the
  Mott and the liquid behaviour as the temperature for which $\kappa$
  departs from its low $T$ value by more than 5\%.
\label{kappa}}
\end{figure}

\begin{figure}[h]
\includegraphics[width=8.5cm]{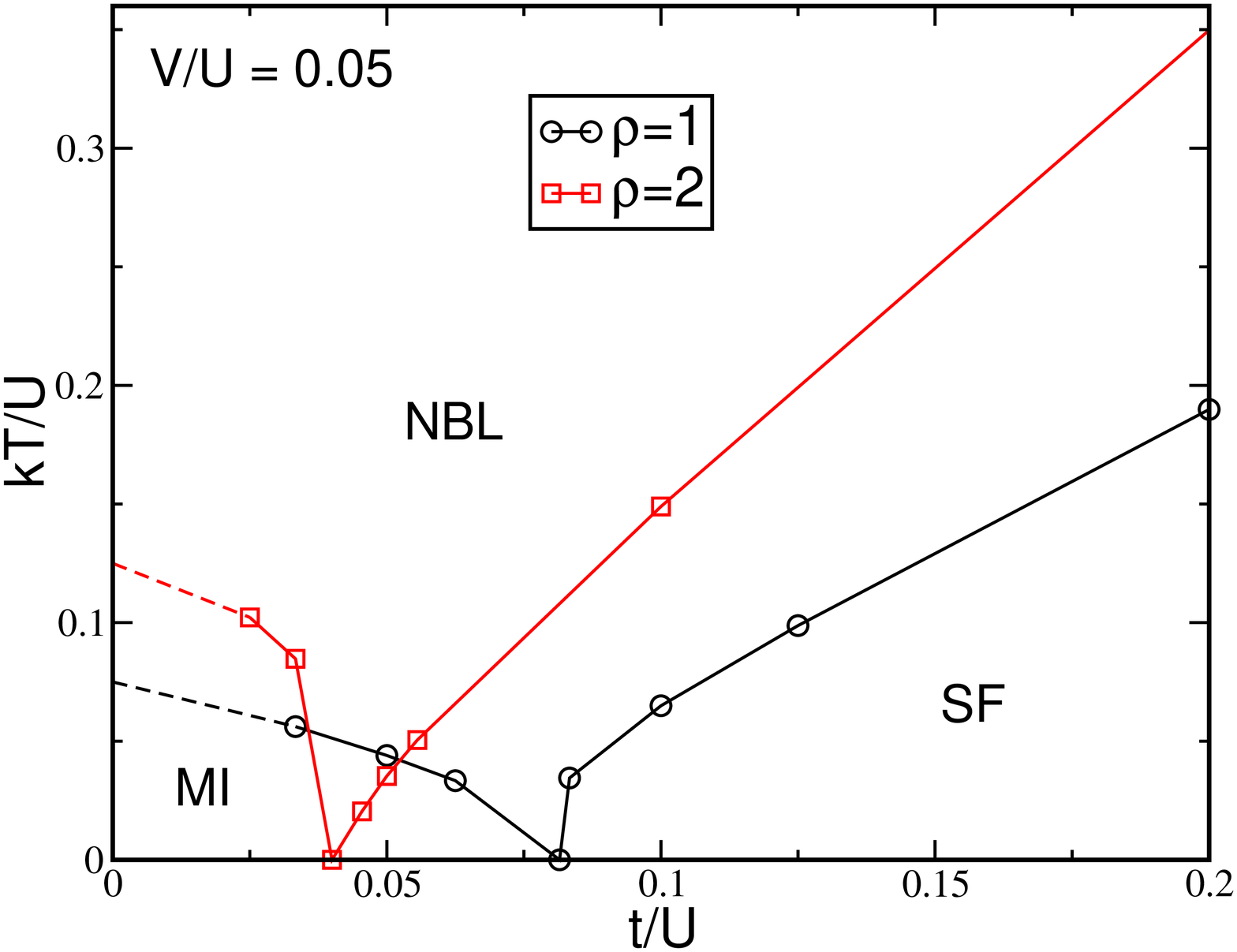}
\caption{The phase diagrams for $\rho=1$ and $\rho=2$ as a function of
  $t/U$ and $kT/U$ at $V/U=0.05$. There are three different regions
  in each diagram: a Mott region (MI), the normal Bose liquid (NBL)
  and the superfluid region (SF). The SF-NBL limit is a BKT
  transition. The MI-NBL limit is a crossover between weakly and
  highly compressible regimes.
\label{diagtemp}}
\end{figure}

Putting these results together, we obtain the phase diagrams shown in
Fig.~\ref{diagtemp} for $\rho=1$ and $\rho=2$. We placed at $T=0$ the
point of the quantum phase transition observed in Sec. \ref{secqpt}.

We now turn to the magnetic behaviour at finite temperature.  In the
Mott phase at $\rho=1$, there is no magnetic order at $T=0$ and this
behaviour persists at finite temperature.  In the Mott phase at
$\rho=2$, the magnetic couplings that lead to a ferromagnetic phase
are weak and the magnetic order disappears for low temperatures $kT
\simeq 0.3$ on an $L=6$ system. As the system is well described, in
the Mott phase, by a Heisenberg model, we do not expect to observe
magnetic order at finite temperature in the thermodynamic limit.

\begin{figure}[h]
\includegraphics[width=8.5cm]{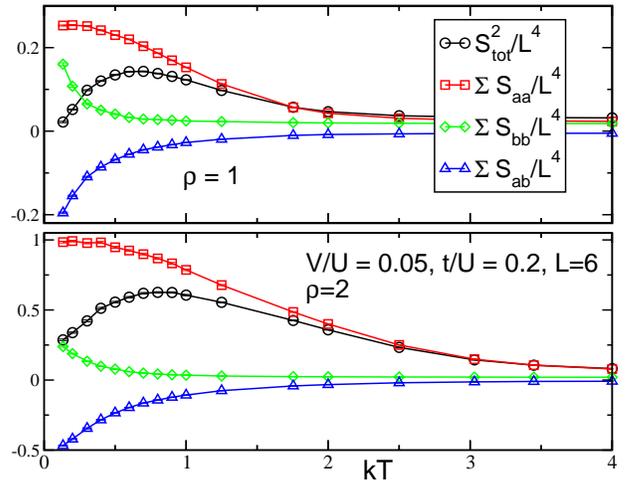}
\caption{Magnetisation of the whole system ($S_{\rm tot}$) and sums
  over all distances of boson-boson ($S_{aa}$), spin-spin ($S_{bb}$)
  and boson-spin $(S_{ab}$) magnetic correlations for $\rho=1$ in the
  superfluid phase. The total magnetisation $S_{\rm tot}$ is zero at
  $T=0$ but rises when $T$ is increased.  As the bosons decouple from
  the spin, when $kT > V$, and no longer form singlets with those,
  they develop ferromagnetic correlations. Top panel: $\rho=1$, bottom
  panel: $\rho=2$.
\label{fig:finiteT_mag_rho12}}
\end{figure}

\begin{figure}[h]
\includegraphics[width=8.5cm]{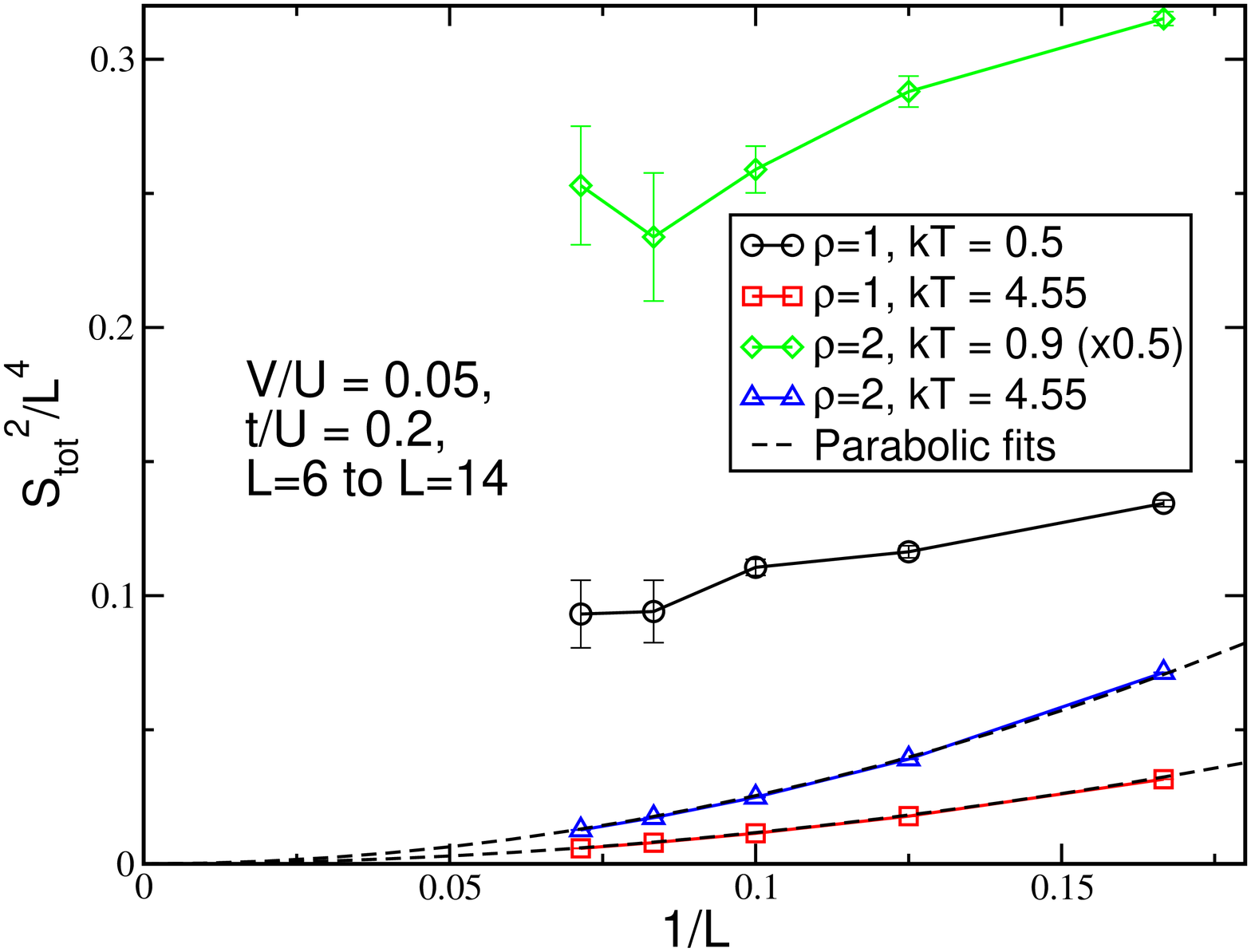}
\caption{Extrapolation of the magnetization as a function of the
  size. At high temperature ($kT = 4.55$) the magnetization
  extrapolates to zero. At the temperatures where we observed the
  peaks of the moment in Fig. \ref{fig:finiteT_mag_rho12} the
  magnetization does not seem to extrapolate to zero. The data for
  $\rho=2, kT = 0.9$ have been divided by 2 for better visibility.
\label{fig:finitesize_finiteT_mag}}
\end{figure}

The behaviour in the superfluid regime is much more interesting.  What
we observe is that the magnetic ordering is reinforced when we
increase the temperature from $T=0$. We find that the total
magnetisation $S_{\rm tot}$ increases before decreasing again and
reaching zero in the high temperature regime (see Fig.
\ref{fig:finiteT_mag_rho12}).  The increase of the total magnetisation
is due to a strong decrease of the boson-spin $S_{ab}$
correlations. This is easily understood as the coupling $V$ between
bosons and spins takes a small value ($V/U=0.05$).  Thermal
excitations break the correlations between spins and bosons. The AF
correlation between those two species disappears and the spins are
then disordered as the bosons no longer mediate an intersite coupling.
As the bosons become independent of the spins, they form
  a FM superfluid with a larger total magnetisation. Once again, the
  mechanism for this FM ordering is not clear but it is obviously
  mediated by the hopping of bosons, the only effect that can couple
  distant particles in this system.  As the hopping $t$ is larger
than $V$ these remaining FM correlations disappear only at larger
temperatures.  For example, for $\rho=1$ (Fig.
\ref{fig:finiteT_mag_rho12}) the AF correlations $S_{ab}$ and the
spin-spin correlations $S_{bb}$ have almost disappeared for $kT = 1$
whereas the FM correlations $S_{aa}$ become negligible only for $kT >
3$.

A non zero magnetisation should not be present at finite temperature
in two dimensions as it contradicts the Mermin-Wagner theorem
\cite{mermin66}. We looked at the evolution of the magnetisation with
sizes for different temperatures
(Fig. \ref{fig:finitesize_finiteT_mag}) to check if it decays to zero.

At high temperature ($kT = 4.55$) the total magnetisation indeed goes
to zero. With no correlations between sites, $S^2_{\rm tot}$ scales as
$L^2$, as the only remaining contributions are on site terms. Hence
the observed behaviour where $S^2_{\rm tot}/L^4 \propto 1/L^2$.
However, at intermediate temperature, close to the maxima of $S^2_{\rm
  tot}$ observed in Fig. \ref{fig:finiteT_mag_rho12}, we do not find a
clear decay of the magnetisation with size. Data obtained for $L=12$
and $L=14$ have large error bars and it is difficult to draw a
conclusion for the thermodynamic limit, but the behaviour does not
seem to correspond to an exponential decay of the magnetic
correlations with distance.

Our interpretation of these data is that, in the superfluid phase, we
have a quasi-long range order of the different Green functions.  We
are then also expecting a quasi-long range order for the magnetic
correlations between the bosons as they are directly related to the
anticorrelated Green functions. This would explain the surprisingly
large values of $S_{\rm tot}^2$ on our small size systems.  A similar
behaviour was found in another spin 1/2 bosons model at finite
temperature \cite{deforges12}.


\section{Conclusion}

We have studied a Bosonic Kondo-Hubbard model with an AF interaction
between spin-$1/2$ bosons and fixed spins. We have drawn the phase
diagram of the system and found that the presence of the Kondo
interaction with the spins facilitates the localisation of the
particles into Mott phases. We have shown that exchange moves are
taking place in Mott phases with density larger than one. 
  Studying the nature of the phase transition, we have always observed
  continuous transitions, contrary to the MF prediction
  \cite{fossfeig11} that a discontinuous transition is present at the
  tip of the $\rho=1$ Mott lobe for low enough $V$.

The magnetic properties of the system are particularly interesting. At
zero temperature, the total magnetization of the system is always
constant but different behaviours can nevertheless be observed. In the
$\rho=1$ Mott phase, we have observed on-site singlets between the
bosons and the spins, with no long range order.  On the contrary, in
the superfluid, we have a FM order of the bosons and the spins and an
AF order between them.  At $\rho=2$ we always have this same magnetic
behaviour for all interactions but with two limiting regimes in the
Mott and superfluid phases.  Notably, we observed the very small value
of the spin-spin correlations at large $U$ that was predicted
analytically.

At finite temperature, we determined the boundary of the superfluid
phase and found the crossover temperature between the Mott and liquid
regions.  More interestingly, we found that, in the superfluid phase,
the total magnetisation is increased due to the fact that the bosons
decouple from the spins. This is unexpected and is certainly not
present in the thermodynamic limit but should be observed on finite
size systems.

The underlying physics of the boson Kondo Hamiltonian studied here has
significant similarities, but also several differences, from the
fermionic case.  For fermions at commensurate filling ($\rho=1$),
although there is a singlet-antiferromagnetic phase transition, both
magnetic phases are insulating, whereas bosons at weak coupling are
superfluid.  The nature of the magnetic order is also somewhat
different.  In the fermionic case, ordering of the local spins
separated by a distance ${\bf r}$ is mediated by a
Rudermann-Kittel-Kasuya-Yoshida interaction which has a modulation
cos(${\bf k_F} \cdot {\bf r}$) where ${\bf k_F}=(\pi,\pi)$ is the
Fermi wavevector at $\rho=1$.  In contrast, the order in the bosonic
case studied here is ferromagnetic.  There is much past and current
interest in the Kondo Hamiltonian for fermions for various sorts of
dilution and randomness in order to model novel quantum phase
transitions and also chemical doping of heavy fermion materials \cite{wang06,seo14,assaad02}.  It
would be interesting to study analogous effects in the boson-Kondo
Hamiltonian.

\medskip

\acknowledgements 

We would like to thank L. De Forges De Parny, M. Foss-Feig and A-M. Rey for stimulating
discussions.  This work was supported by the CNRS-UC Davis EPOCAL
joint research grant. The work of VGR was supported by NSF Grant No. OISE-0952300.
The work of RTS was supported by the Office of the President of the
University of California.

\end{document}